\documentclass[sigconf]{acmart}
\AtBeginDocument{%
  }

\usepackage{xcolor}
\definecolor{placeholdercolor}{RGB}{255,80,80}

 \usepackage{graphicx} 
 \usepackage{subcaption}
\usepackage{colortbl}  
\usepackage{hyperref}
\usepackage{makecell}
\usepackage[inline]{enumitem} 
\usepackage{framed}

\begin{document}

\title{Breaking Models to Test the Judge: A Mutation Testing Approach for Semantic Evaluators of Domain Class Diagrams}


\author{Kevin Delcourt}
\orcid{0009-0005-2988-7308}
\affiliation{%
  \institution{Université de Montréal}
  \department{DIRO}
  \city{Montreal}
  \country{Canada}
}
\email{kevin.delcourt@umontreal.ca}

\author{Meriem Ben Chaaben}
\orcid{0000-0001-8133-0199}
\affiliation{%
  \institution{Université de Montréal}
  \department{DIRO}
  \city{Montreal}
  \country{Canada}}
\email{meriem.ben.chaaben@umontreal.ca}

\author{Abdelhamid Rouatbi}
\orcid{0009-0007-1617-6827}
\affiliation{%
  \institution{Université de Montréal}
  \department{DIRO}
  \city{Montreal}
  \country{Canada}}
\email{abdelhamid.rouatbi@umontreal.ca}

\author{Luciano Marchezan}
\orcid{0000-0003-3096-580X}
\affiliation{%
  \institution{Université de Montréal}
  \department{DIRO}
  \city{Montreal}
  \country{Canada}
}
\email{lucianomarchezan94@gmail.com}

\author{Houari Sahraoui}
\orcid{0000-0001-6304-9926}
\affiliation{%
  \institution{Université de Montréal}
  \department{DIRO}
  \city{Montreal}
  \country{Canada}
}
\email{sahraouh@iro.umontreal.ca}

\renewcommand{\shortauthors}{Delcourt et al.}

\begin{abstract}
In software engineering, many semantic modeling tasks lack a unique ground truth, as human judgments are both costly and subjective. 
This paper explores mutation testing as a scalable alternative for evaluating semantic judges (e.g., LLM-based) of models.
We propose a mutation testing approach in which controlled semantic defects are injected into domain class diagrams.
Starting from pairs of PlantUML class diagrams and textual system descriptions, we apply mutation operators (e.g., removing a class) to generate faulty variants. A candidate judge is then evaluated based on its ability to detect the injected defects.
We define 11 mutation operators for the task of comparing a domain class diagram against a textual description and evaluate the proposed approach against a conventional manual assessment of judgment validity. 
Across six judge configurations (three LLMs and two prompt variants), the automated mutation testing approach is largely consistent with the manual assessment in identifying the better-performing configurations.
The results suggest that mutation testing may serve as a scalable proxy for analyzing semantic judges.
\end{abstract}


\copyrightyear{2026}
\acmYear{2026}
\setcopyright{cc}
\setcctype{by}
\acmConference[MODELS 2026]{ACM/IEEE 29th International Conference on Model Driven Engineering Languages and Systems}{October 04--09, 2026}{Málaga, Spain}
\acmBooktitle{ACM/IEEE 29th International Conference on Model Driven Engineering Languages and Systems (MODELS 2026), October 04--09, 2026, Málaga, Spain}
\acmDOI{10.1145/3822455.3838768}
\acmISBN{979-8-4007-2809-9/2026/10}


\maketitle

\section{Introduction}\label{sec:intro}
Models are widely used in software engineering, and in many model-based approaches they serve as primary artifacts for describing, analyzing, and communicating aspects of software systems~\cite{brambilla2017model}. 
Domain-level class diagrams are important as they capture the main concepts of an application domain, their attributes, and relationships. Beyond their descriptive role, they also automate tasks such as model transformation and generation, among others~\cite{burgueo2025automation}.

A recurring challenge is assessing whether a model correctly reflects an informal or semi-formal textual description~\cite{bucchiarone2020grand}. This is especially an issue when models must be assessed against natural-language requirements, which may be ambiguous or expressed at different levels of abstraction~\cite{chaaben2024Utility, camara2023assessment}. 
Moreover, for many modeling tasks, there is no unique correct solution as different diagrams may satisfy the same description while making different yet logical choices \cite{liebel2024human}. 
These choices are often influenced by modeling conventions and by the context in which the model is used. As a result, evaluating a model by comparison to one reference model provides only a partial view of its correctness.

Large Language Models (LLMs) are used for natural language understanding, abstraction, and cross-artifact reasoning~\cite{zhao2026survey}, making them attractive for model engineering tasks involving textual descriptions~\cite{chen2023automated}.
In particular, LLMs are starting to be used as judges to assess whether a domain model is semantically adequate with respect to a textual specification \cite{capuano2022reverse}. 
However, using LLMs as model judges introduces a second evaluation problem: how can we determine whether the judgment produced by an LLM is itself reliable? LLMs are inherently non-deterministic, sensitive to prompting choices, and may produce judgments that appear plausible while overlooking important modeling defects~\cite{muttillo2025coupling}.

Mutation testing offers a promising direction to address this problem.
The technique evaluates a validation mechanism by introducing small, controlled defects into an artifact and checking whether they are detected~\cite{bockisch2024mutation}.  
It has been applied in MDE contexts, for instance, to assess the adequacy of test suites for model transformations~\cite{mottu2006mutation}, 
or 
for search-based model-driven engineering~\cite{struber2017generating}.
However, existing approaches target the evaluation of rule-based validation oracles,
rather than the evaluation of semantic judges. 

In this paper, we adapt this principle to evaluate LLMs used as semantic judges of domain class diagrams. 
We start from a model--description pair and introduce controlled defects through model transformation (e.g., by removing a class). 
The intuition is that, although it may be difficult to define a unique correct model for a textual description, it is easier to define changes that make a given model less adequate. 
These mutations create model variants whose defects are known by construction. 
An LLM judge can then be evaluated according to whether it detects or penalizes the defect introduced by the mutation. 
The behavior of the LLM judge on these mutants provides evidence of its sensitivity to different categories of modeling defects.

In summary, we make the following contributions:
\begin{enumerate*}[label=(\roman*), itemjoin={{; }}, itemjoin*={{; and }}]
\item We propose a mutation-testing-based approach for validating semantic judges of domain-level class diagrams 
\item We construct a mutated dataset of domain-level class diagrams by applying mutation operators that target classes, attributes, and associations to an existing curated dataset~\cite{verbruggen2025toward} 
\item We empirically show how the mutation-testing-based approach can be used to analyze and compare different LLMs and prompting strategies by comparing our approach to a standard manual evaluation. 
\end{enumerate*}


\section{Background and Related Work}

\subsection{Judging Semantic Tasks in Modeling}

We use the term \emph{semantic task} to denote a task in which success is measured primarily by the semantic quality of the output. 
Examples include many natural-language tasks~\cite{agirre2014semeval}, such as text summarization~\cite{zhang2025systematic}, and emotion recognition in text~\cite{zhang2024deep}.
In MDE, typical semantic tasks include comparing a domain model with a natural-language requirement~\cite{yang2022towards}, or assessing whether two alternative models make equivalent domain commitments~\cite{netz2024natural}. For example, given a requirement about patients booking appointments with doctors, one diagram may explicitly model a \texttt{Receptionist} as an actor, whereas another may abstract this into a generic \texttt{Booking} concept. Which is preferable depends on the modeling purpose, stakeholder priorities, and downstream use. This creates an oracle problem. For many semantic modeling tasks, there is no single reference solution, and element overlap or syntactic well-formedness does not determine whether a model is the most appropriate interpretation of an informal requirement~\cite{nelson2012conceptual,hachm2025towards}.

\subsection{Validating Semantic Judges}

LLM-based semantic judges are increasingly used to evaluate open-ended modeling artifacts using rubrics, contextual information, and natural-language explanations~\cite{bavaresco2025llms,gu2024survey}.
However, these judges inherit the same oracle problem. If the target quality is contextual or partly subjective, then the judge's assessment cannot be validated by comparison with a unique ground truth.

Semantic judges are commonly validated by comparing their assessments against human judgments. Benchmarks such as MT-Bench~\cite{bai2024mt}
measure agreement with human preferences and report that strong LLM judges can approximate human judgments in several settings~\cite{zheng2023judging}.
However, human evaluation is costly, difficult to scale, and subjective, making human agreement alone an incomplete validation criterion, especially for model-related tasks~\cite{pan2024human}. Other approaches explore the use of LLM internal signals, in particular perplexity, as metrics for assessing output quality; that is, the more uncertain a model is about its answer, the more likely the answer is to be of low quality. 
For text-writing tasks, \citet{murugadoss2025evaluating} compared a perplexity-based evaluation method against human judgments and found a weak-to-moderate correlation. 
While promising, this kind of approach is limited because it is (i) not task-specific, and therefore needs to be validated for each new task, and (ii) specific to LLMs, making it difficult or impossible to apply to closed-source models. Finally, a natural extension is to use a jury of heterogeneous judges and aggregate their decisions, while treating inter-judge disagreement as an uncertainty signal. 
For example, The PoLL approach shows that panels of diverse models can reduce single-model bias~\cite{verga2024replacing}, but multiple judges may still share correlated blind spots if trained on similar data~\cite{zheng2023judging,shi2025judging,ye2025justice}.

\subsection{Mutation testing}
Mutation testing evaluates a validation oracle by introducing small, systematic changes into an artifact, producing \emph{mutants}~\cite{ bockisch2024mutation}. Each change is generated by a \emph{mutation operator}. If the oracle detects the difference, the mutant is \emph{killed}; otherwise, it \emph{survives}. The aggregate metric is the \emph{kill ratio}: the proportion of mutants killed. Surviving mutants reveal specific fault classes not detected by the oracle.

Although mutation testing is most commonly associated with assessing test-suite adequacy, its underlying principle has been applied in other fields \cite{john2026mutation,semerath2020diversity}.
In classic test-suite validation, approaches such as LLMorpheus prompt an LLM to generate context-aware mutants that better resemble realistic faults~\cite{tip2025llmorpheus}. Similar approaches are being explored in MDE, for example to generate mutants of Simulink-Stateflow models~\cite{valle2026exploring}, showing that LLM-generated mutants are, on average, more diverse and take less time to generate than hand-made mutants.
Another emerging direction uses mutation-style perturbations to evaluate LLMs and LLM-based judges.
For instance, Crupi et al. use mutation-style perturbations to evaluate whether LLM judges can distinguish incorrect code from behavior-preserving variants~\cite{crupi2025effectiveness}.

In MDE, mutation testing has also been applied directly to modeling artifacts and model transformations.
Mutation operators have been used to emulate realistic transformation faults and to guide the generation of mutant-killing rules~\cite{troya2015towards,guerra2019towards}.
Mutation has also been applied at the model level, for example through mutation operators for UML class diagrams~\cite{granda2016mutation}, 
or language-independent frameworks such as Wodel-Test~\cite{gomez2021wodel}.
However, these works evaluate rule-based validation oracles. The closest work to ours is Ahmed et al.~\cite{ahmed2025mcet}, who manually introduced design smells into sequence diagram/textual requirement pairs and measured how various LLMs detected these issues, demonstrating the applicability of a mutation-like approach to LLM judge validation.



\section{Mutation Testing for Semantic Tasks}

\subsection{Overview}

\begin{figure*}
    \centering
    \vspace{-.3cm}
    \includegraphics[width=\linewidth]{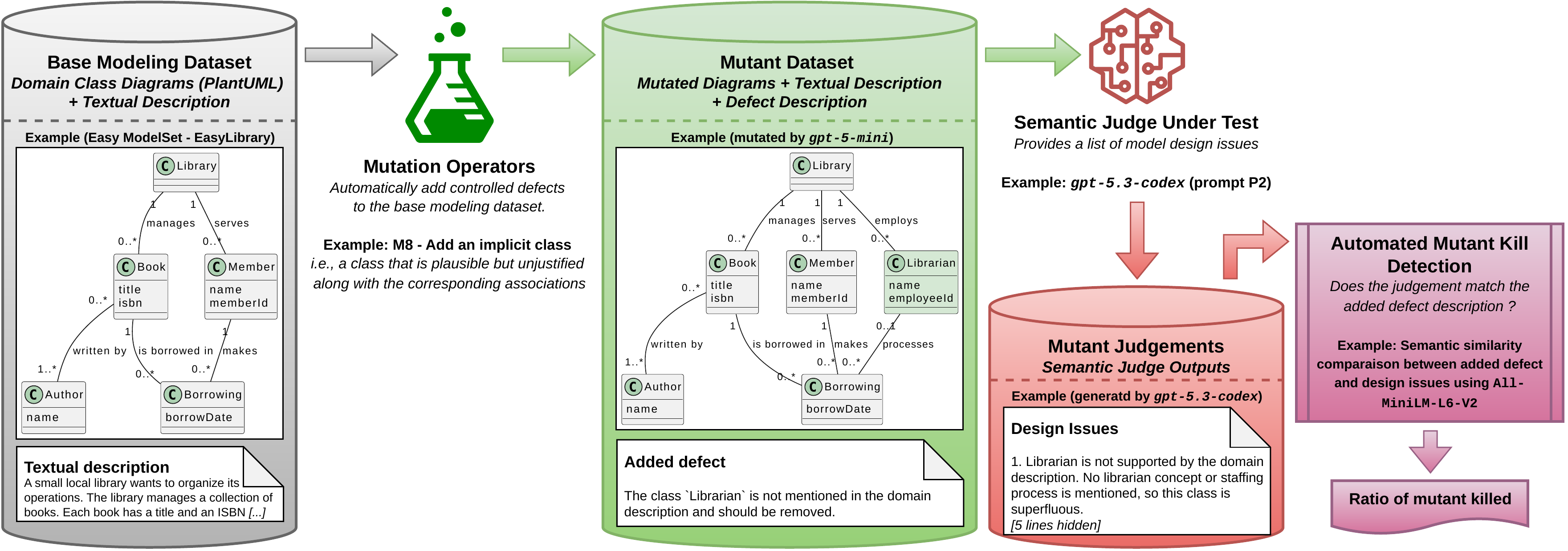}
    \vspace{-.6cm}
    \caption{Mutation-testing-based approach for the validation of semantic judges.}
    \label{fig:conceptual-framework}
    \vspace{-.3cm}
\end{figure*}

Figure~\ref{fig:conceptual-framework} summarizes the proposed approach.
The approach starts from a base dataset of model--description pairs, where each entry consists of a PlantUML domain class diagram and a textual description of the system being modeled. 
Mutation operators introduce known semantic defects into diagrams. For each generated mutant, the operator produces the mutated PlantUML diagram and a concise natural-language description of the injected defect. 
This defect description acts as a lightweight oracle for the evaluation of the judge. 
The semantic judge under test then reports design issues concerning the adequacy between the textual description and the mutated class diagram. 
A mutant is considered killed when the judgment identifies the injected defect. 
This decision is automated through a kill detection strategy that compares the defect description with the judge's reported issues, which in our case is achieved through sentence-embedding similarity. 
The final outcome is the ratio of killed mutants, which provides a quantitative indication of the judge's ability to detect semantic defects.

\subsection{Application to Domain Class Diagram}\label{sec:application}

Here, we describe the specific elements that need to be defined  to apply our approach to domain class diagrams. To transfer this approach to new tasks, the following elements need to be adapted:

\textit{\textbf{Base Modeling Dataset.}}
The base dataset provides the original inputs to which mutations are applied.
It does not need to contain exhaustive annotations or gold-standard judgments. However, the approach assumes that the base dataset contains entries that are sufficiently coherent for the injected mutations to introduce identifiable defects.
For example, in the task of checking a class diagram against a textual specification, if a low-quality class diagram in the base dataset already contains a superfluous class, then removing that class removes a defect instead of adding one. Therefore, the quality of the dataset must be relatively good. We discuss this limitation further in Section~\ref{sec:discussion}.

\textit{\textbf{Mutation Operators.}}
We define 11 mutation operators for domain class diagrams, listed in Table~\ref{tab:mutations}. Each operator injects a semantic defect into the class diagram while keeping the textual description unchanged. All mutations are applied programmatically by parsing the PlantUML source, modifying the relevant elements according to model transformation principles, and producing a new PlantUML file. 
Each operator also generates a concise defect description, such as ``the class Library is missing'', which serves as the lightweight oracle for the kill detection step. See the replication package~\cite{zenododo} for additional operator implementation details.

The operators are divided into two categories. \textit{Automated operators} (M1--M5) are rule-based model transformations that select a target element at random and apply a fixed modification, such as removal or reversal. These can create, at low cost, semantic defects that a competent judge should detect. 
\textit{LLM-assisted operators} (M6--M11) are used when the mutation requires generating a semantically meaningful model element. We use LLM assistance for these operators because the generated element must be realistic enough to make the mutant challenging to detect, while still constituting a defect under the modeling requirements adopted in this study. 

This set of 11 operators 
corresponds to defects that are meaningful in the context of domain class diagrams.
However, the main challenge in evaluating semantic modeling tasks is that there is no single ground truth and, as such, no commonly accepted set of rules for defining what constitutes a good model~\cite{liebel2024human}. 
Consequently, the operators should be adapted each time the approach is applied in a new context.

\newcolumntype{Y}{>{\raggedright\arraybackslash}X}
\begin{table}[]
\centering
\scriptsize
\setlength{\tabcolsep}{2.5pt}
\renewcommand{\arraystretch}{0.92}
\caption{Mutation operators for domain class diagrams.}
\label{tab:mutations}
\vspace{-.25cm}
\begin{tabular}{@{}c c p{0.31\columnwidth} p{0.34\columnwidth}@{}}
\toprule
\textbf{ID} & \textbf{Impl.} & \textbf{Transformation} & \textbf{Expected violation} \\ 
\midrule
M1 & Auto & Remove a class. & Required concept is missing. \\
M2 & Auto & Remove an attribute. & Required property is missing. \\
M3 & Auto & Remove a relationship. & Required domain relation is absent. \\
M4 & Auto & Add an irrelevant association with multiplicities. & Unjustified relation or multiplicity appears. \\
M5 & Auto & Reverse a non-symmetrical relationship. & Direction or role semantics contradict the description. \\
\midrule
M6 & LLM & Add an implementation-oriented class. & Technical artefact appears as a domain concept. \\
M7 & LLM & Add an implementation-oriented attribute. & Technical property appears in a domain class. \\
M8 & LLM & Add an implicit domain class. & Plausible but unjustified class is added. \\
M9 & LLM & Add an implicit domain attribute. & Plausible but unjustified property is added. \\
M10 & LLM & Add a class from another domain. & Unrelated concept is introduced. \\
M11 & LLM & Add an attribute from another domain. & Unrelated property is introduced. \\
\bottomrule
\end{tabular}
\vspace{-.3cm}
\end{table}

\textit{\textbf{Semantic Judges.}}
The semantic judges receive mutated entries and produce a judgment for the semantic task at hand, i.e., an assessment of the design issues in domain class diagrams when compared to their respective system descriptions.
Although this work focuses on evaluating LLM-based judges, the approach can also be applied to other semantic validators, including rule-based or manual approaches.
In all cases, the judge is evaluated according to whether it detects the defect intentionally introduced by the mutation operator.

\textit{\textbf{Mutant Kill Detection.}}
Since we aim to automate the evaluation process, we define a step that automatically determines whether the judge identified the introduced mutations. To do so, we compare the injected defect description with each issue reported by the judge using sentence-embedding similarity.

Let $d_m$ be the defect description associated with mutant $m$, and let $\mathcal{I}_m$ be the set of issues reported by the judge for that mutant. We encode both $d_m$ and each issue $i \in \mathcal{I}_m$ using \texttt{all-MiniLM-L6-v2}~\cite{sentence-transformers-all-MiniLM-L6-v2}, a lightweight sentence transformer that offers a good trade-off between encoding quality and computational cost for short textual comparisons.
We then compute the cosine similarity between the resulting embeddings.
If the judge reports no issue, the mutant is considered alive.
A mutant is considered killed when at least one reported issue reaches the similarity threshold $\tau$: \begin{equation}
\mathrm{killed}(m)=1 \Longleftrightarrow \max{(\forall i \in \mathcal{I}_m, \cos(E(d_m), E(i)))} \geq \tau.
\end{equation}

Other strategies may also be possible. Depending on the task, the comparison may rely on semantic similarity, natural-language entailment, keyword matching
or a combination of these techniques.


\section{Evaluation}

\begin{figure*}[t]
    \vspace{-.3cm}
    \centering
    \includegraphics[width=.99\linewidth]{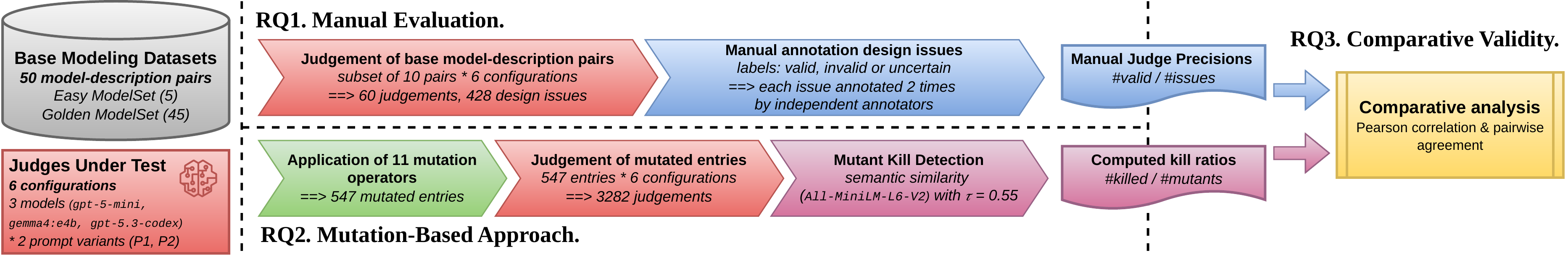}
    \vspace{-.4cm}
    \caption{Overview of the evaluation procedure.}
    \label{fig:evaluation-process}
    \vspace{-.2cm}
\end{figure*}

To validate our mutation-testing-based approach, we compare it with a standard manual validation of LLM-based judges, as illustrated in Figure~\ref{fig:evaluation-process}.  This three-step evaluation process  answers the following research questions:

\begin{leftbar}
\noindent
\textbf{RQ1--Manual evaluation:} How precise are the semantic issues reported by the evaluated judge configurations?

\noindent
\textbf{RQ2--Mutation-based approach:} How effectively do the evaluated judge configurations detect controlled semantic defects introduced through mutation?

\noindent
\textbf{RQ3--Comparative validity:} To what extent can the mutation kill ratio reproduce the comparative conclusions obtained from a manual evaluation?
\end{leftbar}

\textit{\textbf{Judges under test.} }
We evaluate six LLM-based judge configurations, formed by three LLMs (open-source \texttt{gemma4:e4b}, and closed-source \texttt{gpt-5-mini} and \texttt{gpt-5.3-codex}) and two prompt variants (P1 and P2). 
The three LLMs were selected because they have different sizes and capabilities. 
Both prompts are zero-shot prompts that instruct the LLM to provide a list of class-diagram design issues, with P2 providing more detailed guidance than P1. 
The replication package~\cite{zenododo} contains the full prompts. 
These conditions are intended to reflect a standard behavior in LLM-based judge development, where practitioners test different models and prompts to identify the most suitable configuration.

\textit{\textbf{Base Datasets.}} 
Table~\ref{tab:modelsets} presents the two datasets used in this study. For each dataset, size is measured through the number of model elements, i.e., classes, attributes, and associations. 
Lexical diversity is measured from diagram labels, i.e., class names, attribute names, and association labels. These metrics were computed using the CM benchmark for model sets introduced by Glaser et al.~\cite{glaser2026benchmarking}.

\begin{table}[h]
\vspace{-.2cm}
\centering
\footnotesize
\caption{Overview of the datasets used in this study.}
\label{tab:modelsets}
\vspace{-.2cm}
\begin{tabular}{@{}p{0.55\linewidth}cc@{}}

\toprule
\textbf{Metric} & \textbf{Easy} & \textbf{Golden}~\cite{verbruggen2025toward} \\
\midrule
Number of model--description pairs & 5 & 45 \\
\midrule
Avg. elements per model & 24.4 & 54.3 \\
Min. elements per model & 22 & 14 \\
Max. elements per model & 29 & 114 \\
\midrule
Total number of labels & 62 & 1190 \\
Avg. labels per model & 12.4 & 26.4 \\
Total vocabulary size & 50 & 776 \\
Avg. vocabulary per model & 10.0 & 17.2 \\
\bottomrule
\end{tabular}
\vspace{-.2cm}
\end{table}

The \textit{Golden ModelSet}~\cite{verbruggen2025toward} is a collection of expert-curated domain class diagrams paired with textual descriptions. 
It contains 45 PlantUML class diagram and description pairs and provides realistic cases for semantic model validation, made by experts from various universities. 
Its diagrams are relatively complex, with an average of 54.3 elements per model. To complement it with simpler and more controlled examples, we define the \textit{Easy ModelSet}, a small set of five smaller entries manually created by the authors. 
These easy cases are included in the replication package and were used to calibrate the approach on examples where the expected semantic defects are easier to inspect manually.

\subsection{RQ1. Manual Evaluation}
RQ1 estimates the precision of the issues reported by each judge configuration on non-mutated model--description pairs, providing a baseline measure of judgment validity.

\textit{\textbf{Setup.}}
We evaluated the six judge configurations on ten non-mutated entries: all five entries from the Easy ModelSet and five selected at random from the Golden ModelSet. 
Each configuration produced one judgment per pair, for a total of 60 judgments.
Four independent annotators with MDE expertise and at least master's-level training participated in the annotation. 
Each annotator reviewed 30 judgments, such that every judgment was independently reviewed by two annotators. 
At the level of individual reported issues, annotators assigned one of three labels: \emph{valid}, when the issue was supported by the diagram, the textual description, and the modeling criteria adopted in this study; \emph{invalid}, when the issue was unsupported or resulted from an incorrect interpretation; and \emph{uncertain}, for borderline cases.

We assessed inter-annotator agreement using raw percentage agreement and nominal Krippendorff's $\alpha$, which accommodates the partially crossed annotation design~\cite{krippendorff2011computing}. 
All disagreements and uncertain cases were subsequently adjudicated by an additional annotator to obtain a final valid/invalid label for every reported issue. 
Precision is therefore calculated as the ratio of the number of \emph{valid} issues to the total number of reported issues.

\textit{\textbf{Results.}}
The 60 judgments contained 428 reported issues (86/342 for Easy/Golden ModelSet). 
The judges reported more issues for the Golden ModelSet pairs, which may be explained by the fact its entries are larger and more heterogeneous. 
The annotators assigned the same initial label to 65\% of the issues. 
Krippendorff's $\alpha$ was $0.32$, indicating limited agreement before adjudication. 
This result illustrates that deciding whether a natural-language criticism of a domain class diagram is semantically justified remains subjective, even for annotators with relevant modeling expertise. 
The adjudication process was therefore necessary to obtain a stable reference for the precision analysis.
Figure~\ref{fig:manual-precision} reports precision for each combination of model, prompt, and dataset. 
Pooled by model, \texttt{gpt-5.3-codex} achieves the highest precision at 63.7\%, compared with 31.6\% for \texttt{gemma4:e4b} and 35.5\% for \texttt{gpt-5-mini}. 
Pooling by prompt, P2 reaches 48.6\% precision, compared with 26.9\% for P1. 


\subsection{RQ2. Mutation-Testing Evaluation}\label{sec:evaluation-rq2}
RQ2 evaluates whether the judge configurations detect known semantic defects injected through the mutation operators.

\textit{\textbf{Setup.}}
We applied each mutation operator once to each applicable model--description pair. 
Operator applicability depends on the structure of the base diagram: in some cases, the base diagram from the Golden ModelSet does not include an asymmetric relationship, making M5 inapplicable, or does not include attributes, making M2 inapplicable.
After filtering three cases of non-applicable operator--model combinations, the generated mutant dataset contains 547 mutants.
We executed the six judge configurations once on each mutant, producing 3,282 mutant judgments. 
For each mutant, the semantic-similarity kill detector described in Section~\ref{sec:application} determined whether at least one reported issue matched the injected defect.

The similarity threshold was calibrated on a manually labeled set of 209 mutant judgments produced by \texttt{gpt-5-mini} and set to $\tau=0.55$, at which the kill detector reaches an optimal F1 score of 0.90, with a precision of 0.86 and a recall of 0.95. 
Further calibration details are provided in the replication package~\cite{zenododo}. 
The mutation kill ratio we report therefore corresponds to the number of killed mutants divided by the total number of mutants.

\begin{figure}[]
    \centering

    \begin{subfigure}{\linewidth}
        \centering
        \includegraphics[width=\linewidth]{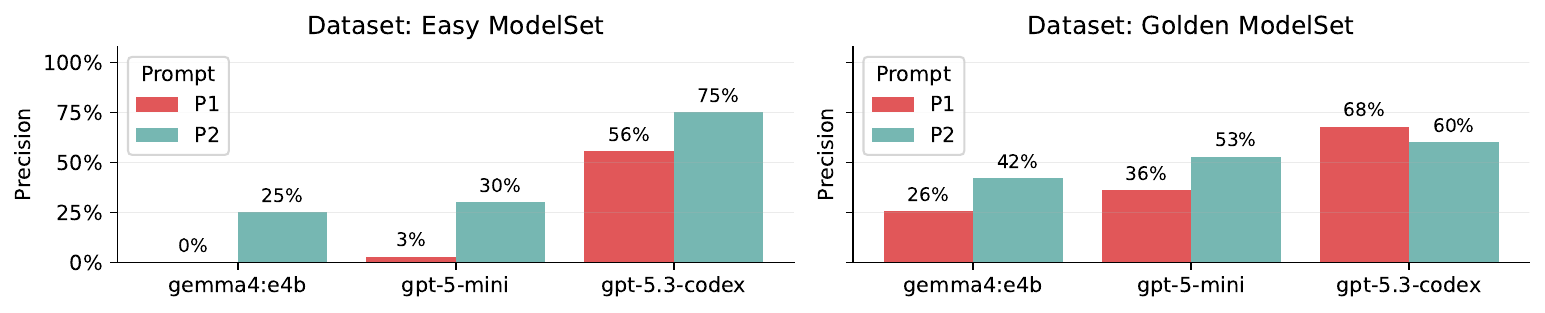}
        \vspace{-.5cm}
        \caption{RQ1: Manual issue-level precision.}
        \label{fig:manual-precision}
    \end{subfigure}

    \begin{subfigure}{\linewidth}
        \centering
        \includegraphics[width=\linewidth]{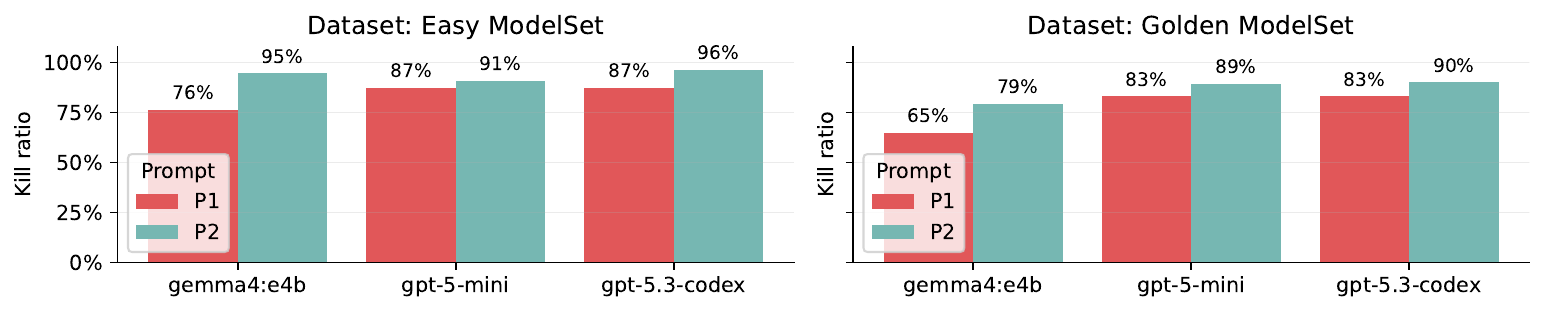}
        \vspace{-.5cm}
        \caption{RQ2: Mutation kill ratios.}
        
        \label{fig:mutation-kill-ratios}
    \end{subfigure}
    \vspace{-.6cm}
    \caption{Results for RQ1 and RQ2 across each LLM, prompt variant, and dataset.}
    \label{fig:rq1-rq2-results}
    \vspace{-.4cm}
    \end{figure}

\textit{\textbf{Results.}}
The overall kill ratio was 82.4\%. 
Figure~\ref{fig:mutation-kill-ratios} presents the results for each combination of model, prompt, and dataset. 
Mutants derived from the Easy ModelSet were killed more frequently than those from the Golden ModelSet, as injected defects may constitute a more evident departure in smaller and more regular diagrams.
Pooled by model, \texttt{gemma4:e4b} obtains the lowest kill ratio at 73.4\%, compared with 86.6\% for \texttt{gpt\allowbreak-5\allowbreak-mini}
and 87.1\% for \texttt{gpt\allowbreak-5.3\allowbreak-codex}.
Pooled by prompt, P2 again outperforms P1, with kill ratios of 87.0\% and 77.7\%, respectively.

\subsection{RQ3. Comparative Evaluation}
\label{sec:evaluation-rq3}

RQ1 and RQ2 measure different and complementary aspects of judge performance. 
RQ3 examines their convergent validity as instruments for comparing judge configurations. 

\textit{\textbf{Setup.}}
For the six judge configurations, we aggregated results across both datasets and constructed one entry for each configuration, containing both the corresponding manual precision and computed mutation kill ratio.
To assess the correlation between the two measures, we computed Pearson's correlation coefficient.  We also computed pairwise ordering agreement, defined as the proportion of configuration pairs for which the two instruments rank one configuration above the other in the same direction.

\normalsize

\textit{\textbf{Results.}}
Across the six conditions, manual precision and mutation kill ratio have a Pearson correlation of $r=0.624$, indicating a moderate-to-strong positive linear association. 
Among the $\binom{6}{2}=15$ condition pairs, the two instruments produce the same ordering for 11 pairs (agreement rate of 73.3\%). 
Some orderings were not conserved. For example, \texttt{gpt-5-mini} frequently detected injected defects (high kill ratio), while also reporting many unsupported issues on non-mutated diagrams (low manual precision).

\section{Discussion}\label{sec:discussion}
\textit{\textbf{Mutation Testing as an Evaluation Instrument.}}
The evaluation highlights that manual annotation and mutation testing are complementary instruments, as a judge may perform well on one dimension without performing well on the other. 
This is particularly important for verbose judges,  as reporting  many possible issues increases the probability of matching an injected defect and can therefore increase the kill ratio, even when many of the additional issues are invalid. 
This behavior was observed in RQ1, especially in the \emph{Easy ModelSet} results.

Nevertheless, RQ3 shows that mutation testing captures a useful comparative signal, leading to the same conclusions regarding prompt and LLM performance.
Because mutant generation and kill detection are largely automated
mutation testing is suited to preliminary comparisons and targeted diagnosis of known defect categories. 
Manual evaluation remains necessary to assess false positives, ambiguous judgments, and aspects of model quality not covered by the mutation operators.

\textit{\textbf{Diagnostic Value of the Mutation Operators.}}
Aggregate kill ratios conceal substantial differences among defect categories. 
Table~\ref{tab:kill-ratio-by-operator}  complements the overall score with results for each mutation operator. 
M10 and M11, which add unrelated elements, are detected particularly frequently. 
These operators remain useful as sanity checks, but
if such easy mutants dominate the aggregate score, they can make judge performance appear stronger than it is on subtler modeling defects. 
They could therefore be retained as a separate basic-capability category, or omitted from repeated experiments when computational cost is the primary concern.

\begin{table}[h]
\vspace{-.1cm}
\centering
\scriptsize
\setlength{\tabcolsep}{4pt}
\renewcommand{\arraystretch}{0.95}

\caption{Overall kill ratio by mutation operator.}
\vspace{-.3cm}
\label{tab:kill-ratio-by-operator}
\begin{tabular}{@{}lccccccccccc@{}}
\toprule
\textbf{Operator} & M1 & M2 & M3 & M4 & M5 & M6 & M7 & M8 & M9 & M10 & M11 \\
\midrule
\textbf{Kill ratio (\%)} &
72.6 & 62.3 & 72.8 & 77.5 & 76.3 & 86.6 & 92.6 & 76.5 & 77.2 & 95.8 & 97.8 \\
\bottomrule
\end{tabular}
\vspace{-.1cm}
\end{table}

In contrast, operators with lower kill ratios identify concrete blind spots in the judge. A low score for relationship-orientation mutations, for example, suggests that the judge does not reason reliably about association semantics. 
This diagnostic information is difficult to obtain from a single aggregate comparison with human judgments, and can directly inform prompt refinement or the definition of additional task-specific examples.

The mutation operators also make the evaluation requirements explicit. 
Their definition encodes the modeling experts' assumptions about completeness, abstraction, relevance, and relationship semantics. 
This is an advantage when the intended criteria are explicit, but it also means that the resulting score is only meaningful with respect to the selected operators and their distribution. 

\textit{\textbf{Threats to Validity.}}
The RQ1 reference is based on 60 judgments over ten baseline entries. 
Only five of the 45 Golden ModelSet entries could be annotated due to limited capacity. 
These five were selected at random to limit selection bias, but a larger sample could change the observed precision values and the relative ordering of configurations. 
Moreover, manual annotation evaluates only reported issues and therefore 
doesn't provide
an exhaustive measure of semantic correctness. 
The initial Krippendorff's $\alpha=0.32$ also shows that the labels are sensitive to subjective interpretations. 
Adjudication resolves disagreements for analysis purposes, but does not eliminate the underlying ambiguity of the task.

Mutation testing assumes that transformations degrade an acceptable model--description pair. 
This assumption can fail when a baseline model is of poor quality, e.g., already contains superfluous elements. 
We manually inspected 79 difficult cases identified during pilot runs and found that 69 were confirmed as valid degradations, while 10 were borderline mutants for which the transformation had little negative effect.
These mutants were retained, which may underestimate the kill ratio of a competent judge. 

The RQ2 results depend on the semantic-similarity detector described in Section~\ref{sec:evaluation-rq2}. 
Although the selected threshold achieves an F1 score of 0.90 on the calibration data, semantic similarity can still produce both false matches and missed paraphrases. 
The threshold was calibrated using judgments from \texttt{gpt-5-mini}; differences in wording across models or prompts may therefore affect transferability. 
A larger, independently held-out test set and comparisons with alternative embedding-based, entailment-based, structured-matching, or LLM-based detectors would provide stronger evidence for the validity of the kill decisions.


\textit{\textbf{Limitations and Research Opportunities.}} Our study represents an initial investigation of mutation testing as an evaluation instrument for LLM-based semantic judges. 
The mutation operators target semantic defects in PlantUML domain class diagrams. Although many of the underlying concepts (e.g., completeness, relevance, relationship semantics, and abstraction) are common across conceptual modeling languages, we do not have evidence on how the approach generalizes to other modeling notations, DSLs, or different semantic tasks. 
Thus, future research can investigate \textit{how mutation-testing performs as an evaluation approach for semantic tasks on other modeling languages and domains}. 
Furthermore, although the choices of prompts and LLMs were sufficient to demonstrate the feasibility of the approach and to compare to manual analysis (RQ3), \textit{additional LLMs and prompts can be tested to explore the robustness of the observed trends} from our results. 

Considering the mutations, our injection strategy applies a single semantic defect to each mutant. This design 
is limited regarding the complexity of real-world models, which often contain multiple interacting inconsistencies. This leads to opportunities to \textit{investigate richer mutation strategies that introduce combinations of defects with varying levels of subtlety and interaction}. 
For example, one can adapt previously proposed UML mutation operators by ensuring they inject semantic defects in UML diagrams beyond class. Similarly, testing our approach with larger diagrams (thousands of elements) is challenging, as guaranteeing that the injected defects break the semantics is harder. 
More importantly, detecting the injected defects may be difficult for LLMs as larger models will lead to a larger context to be analyzed. Hence, future directions can \textit{investigate how to optimize the mutant injection in larger models}, as well as \textit{how to pre-filter the context that the LLMs need to analyze to detect the defects}, concerning the scalability of the mutation-based principle.

\section{Conclusion}

This work proposes a mutation-testing approach for evaluating LLM-based semantic judges of domain class diagrams against textual descriptions. We define a set of domain-class-diagram mutation operators, generate controlled semantic defects, and compare the usefulness of the approach to manually measured judge precision across multiple models and prompt variants.
The results support mutation testing as a practical mechanism for the iterative development of semantic model judges.

Once mutation operators and a kill detection strategy have been established, new models or prompts can be evaluated automatically against the same defect families.
This makes it possible to track regressions, identify capability gaps, and refine prompts before investing in a larger expert annotation campaign.
The approach is particularly attractive in MDE, where experts' time is scarce and semantic tasks do not admit a unique ground truth. More broadly, our objective is to facilitate the engineering of reliable judges for semantic MDE tasks.
Reliable semantic judges could then support downstream MDE activities such as model repair, interactive modeling assistance, quality control for generated-model datasets, and the validation of AI-based modeling tools.

\begin{acks}

This work was partially funded by the Natural Sciences and Engineering Research Council of Canada (NSERC), grant number RGPIN-2025-05677.

\end{acks}

\bibliographystyle{ACM-Reference-Format}
\clearpage 
\bibliography{bib}

\end{document}